\documentclass[conference]{IEEEtran}
\IEEEoverridecommandlockouts
\usepackage{cite}
\usepackage{amsmath,amssymb,amsfonts}
\usepackage{graphicx}
\usepackage{textcomp}
\usepackage{xcolor}
\usepackage{algpseudocode}
\usepackage{algorithm,algpseudocode}
\usepackage{algorithmicx}

\usepackage[normalem]{ulem}

\usepackage{optidef}

\usepackage{amsmath,epsfig,cite,amsfonts,amssymb,psfrag,color}
\usepackage{epstopdf}
\usepackage{subfig}

\def\BibTeX{{\rm B\kern-.05em{\sc i\kern-.025em b}\kern-.08em
    T\kern-.1667em\lower.7ex\hbox{E}\kern-.125emX}}

\def\Uset{\mathcal{U}}
\def\ksum{\sum\limits_{k \in \mathcal{U}}}
\def\jsum{\sum\limits_{j \in \mathcal{U}}}
\def\isum{\sum\limits_{i \in \mathcal{U}}}
\def\jNotKsum{\sum\limits_{j\neq k }^{U}}

\def\kMOnesum{\sum\limits_{j=1 }^{k-1}}
\def\pr{\mathcal{P}}
\def\omegaR{\omega_{k}^R}
\def\omegaN{\omega_{k}^N}

\newcommand{\abss}[1]{{\lvert{#1}\rvert}^2}

\begin{document}
\raggedbottom

\title{Resource Allocation and Performance Analysis of Hybrid RSMA-NOMA in the Downlink}

\author{Mohammad Amin Saeidi, {\em Graduate Student Member IEEE}, and Hina~Tabassum, {\em Senior Member IEEE} \thanks{This research was supported by a Discovery Grant funded by the Natural Sciences and Engineering Research Council of Canada.}
\\
 Department of Electrical Engineering and Computer Science, York University, Toronto, ON, Canada
\\
Email: amin96a@yorku.ca, hinat@yorku.ca
}

\maketitle

\begin{abstract}
Rate splitting multiple access (RSMA) and non-orthogonal multiple access (NOMA) are the key enabling multiple access techniques to enable massive connectivity. However, it is unclear whether RSMA would consistently outperform NOMA from a system sum-rate perspective, users' fairness, as well as convergence and feasibility of the resource allocation solutions. This paper investigates the weighted sum-rate maximization problem to optimize power and rate allocations in a hybrid RSMA-NOMA network. 
In the hybrid RSMA-NOMA, by optimally allocating the maximum power budget to each scheme, the BS operates on NOMA and RSMA in two orthogonal channels, allowing users to simultaneously receive signals on both RSMA and NOMA. Based on the successive convex approximation (SCA) approach, we jointly optimize the power allocation of users in NOMA and RSMA, the rate allocation of users in RSMA, and the power budget allocation for NOMA and RSMA considering successive interference cancellation (SIC) constraints. 
Numerical results demonstrate the trade-offs that hybrid RSMA-NOMA access offers in terms of system sum rate, fairness, convergence, and feasibility of the solutions.

\end{abstract}


\section{Introduction}

Non-orthogonal multiple access (NOMA) and rate splitting multiple access (RSMA) are becoming critical to support multiple users over the same frequency and time resources while leveraging successive interference cancellation (SIC) \cite{9695364}. While the performance of orthogonal multiple access (OMA), NOMA, spatial division multiple access (SDMA)  has been thoroughly investigated in a variety of settings, RSMA has been recently considered as a more generalized, reliable, and efficient multiple access scheme. In contrast to SDMA, which depends on treating any residual interference as noise, and NOMA, which depends on fully decoding interference, RSMA has the capacity of partially decoding the interference and partially treating the interference as noise  \cite{mao2018rate}.

In RSMA, a user's message is divided into a private and a common part. After combining the common messages of all users in one message, the base station (BS) broadcasts the private and common message signals to users. The common signal can be decoded by all users, while each user only needs to decode its corresponding private signal. Utilizing the SIC, the common signal is removed, and then the user can decode the private signal while treating other users' private signals as noise. On the contrary, NOMA is a special case of RSMA, where superposition coding is applied at the BS in the downlink transmission, and the BS transmits the messages of all users using the same time-frequency resource block. By taking advantage of SIC, each user decodes the message of all weaker channel users' and then decodes its own signal.

The work in \cite{SICforNOMA} formulates system sum rate maximization via power allocation in NOMA while taking practical SIC constraints into account.
In \cite{SICforRSMA}, a sub-optimal solution is proposed to obtain private and common power and rate allocations to maximize the system sum-rate. The authors show that the RSMA scheme outperforms NOMA and SDMA. The work in \cite{RSMA-EE-SWIPT-DL} investigates an energy-efficiency maximization problem in a downlink simultaneous wireless information and
power transfer aided RSMA system and proposed a fractional programming based algorithm. 
In \cite{li2020resource-multiCarrier}, the authors investigate a multicarrier RSMA system and solve a sum-rate maximization problem, and the authors in \cite{RSMA-DL-UAV} studied RSMA in a multi users UAV aided downlink system.

The existing research showed that RSMA can consistently outperform NOMA. However, to our best knowledge, there is no research work  that demonstrates the performance of a hybrid RSMA-NOMA scheme which benefits from the features of both RSMA and NOMA. Subsequently, this paper attempts to answer "\textit{whether adopting RSMA would consistently outperform NOMA from system sum-rate perspective as well as fairness perspective? If not, than whether a hybrid NOMA-RSMA access be beneficial?}".

In this paper, for the first time, we formulate and solve a weighted sum-rate maximization problem to jointly optimize the downlink power and rate allocations  considering hybrid RSMA-NOMA. The weights denote the priority of each user. {In the hybrid RSMA-NOMA, by optimally allocating the maximum power budget to each scheme, the BS operates on NOMA and RSMA in two orthogonal channels, allowing users to simultaneously receive signals on both RSMA and NOMA.} Based on successive convex approximation (SCA), we optimize the power allocation of users in NOMA and RSMA channels, rate allocation of users in RSMA  as well as the joint power budget allocation for NOMA and RSMA. Some essential practical constraints are taken into account, including SIC constraints with detection threshold and the power budget allocation between the two schemes. Numerical results demonstrate the performance of the hybrid access in two scenarios: \textbf{ (i)} equal users' weights leading to conventional sum-rate maximization, and \textbf{(ii)} unequal users' weights. For equal users' weights, the hybrid access outperforms NOMA and RSMA in terms of data rate and fairness, respectively. For unequal users' weights, hybrid access outperforms NOMA and RSMA in terms of system sum-rate, and fairness improves compared to the equal weights.

\textit{Organization:} The rest of the paper is organized as follows. Section~II introduces the considered system model. In Section~III, the weighted system sum-rate maximization problem is formulated. Section~IV provides an SCA based solution to the optimization problem. Numerical results are discussed in Section~V, and Section~VI concludes the paper.

\begin{figure}[!t]
   \centering
\includegraphics[scale=0.25]{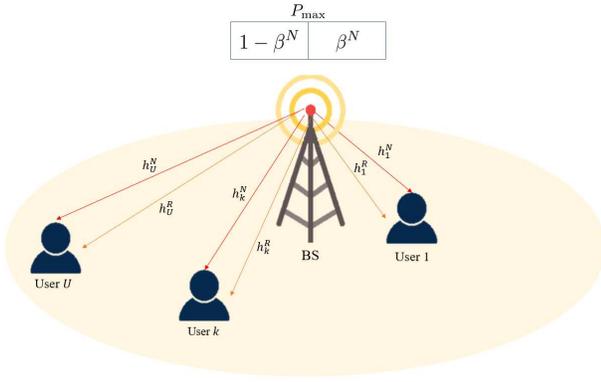}
\caption{A downlink system for hybrid RSMA-NOMA}
\label{fig:SysModel}
\end{figure}

\section{System Model and Assumptions}

We consider a  single antenna BS and a multi-user system consisting of $U$ single antenna users  and denote the set of users as $\mathcal{U}=\{1,\cdots,U\}$.  We consider two orthogonal sub-channels operating on NOMA and RSMA  so they do not experience interference from each other, as shown in Figure \ref{fig:SysModel}. Each user is capable of operating on both the channels. {The BS power budget allocation between NOMA and RSMA channels $\beta$ is optimized such that the BS allocates $\beta P_{\max}$ to the NOMA channel and $(1-\beta)P_{\max}$ to RSMA channel, where $P_{\max}$ denotes the maximum power budget at the BS. Based on the system demand and users' positions, BS can effectively allocate a proportional power budget to each of the NOMA and RSMA channels.} The received signals and data rate of users in NOMA and RSMA channels can be modeled, respectively, as:
\subsubsection{NOMA Channel} The received signal at user $k$ in NOMA can be written as follows:
\begin{equation}
y_k^N = h_k^N \sqrt{p_k^N} s_k + h_k^N \kMOnesum \sqrt{p_j^N} s_j + n,
\end{equation}
where $h^N_k \in \mathbb{C}$ is the channel coefficient between the BS and user $k$, where superscript $N$ denotes the sub-channel used in NOMA, $s_k$ is the symbol intended to user $k$, $p_k^N$ is the transmit power for user $k$ at the BS in NOMA, and $n$ denotes the additive white Gaussian noise (AWGN) with zero mean and $\sigma^2$ as its variance. Then the achievable rate of user $k$ is expressed as:
\begin{equation}\label{eq:NOMA_rate}
    R_k^N = \log_2(1 + \gamma_k^N) = \log_2\left(1+\frac{\abss{h^N_k}p_k^N}{\abss{h^N_k}\kMOnesum p_j^N + \sigma^2}\right),
\end{equation}
where the signal to noise and interference ratio (SINR) of user $k$ in NOMA is denoted by $\gamma_k^N$. Applying SIC, user $k$ subtracts off users $k+1$ to $U$ signals from its received signal and then decodes its message. Thus, only the signals of remaining users with stronger channels are considered as \textcolor{black}{noise}, i.e., $\abss{h^N_k} \kMOnesum p_j^N$. 
It is assumed that channel gains of users are ranked in descending order $\abss{h_1}\geq\abss{h_2}\geq \cdots \geq\abss{h_U}$ \cite{zeng2019energy-Channel-gain}. Thus, the necessary condition for efficient SIC can be stated as follows:
\begin{equation}\label{eqn:NOMA}
    p_k^N \delta^N_{k-1} - \kMOnesum p_j^N \delta^N_{k-1}\geq P_{\mathrm{tol}} , k=2,...,U,
\end{equation}
which is referred as \textit{SIC constraint} \cite{SICforNOMA}, $\delta^N_k = \frac{\abss{h^N_k}}{\sigma^2}$, and $P_{\mathrm{tol}}$ is the minimum power difference required to differentiate the signal to be decoded and the rest of the non-decoded signals. 

\subsubsection{RSMA Channel}  The received signal at user $k$ in RSMA channel is given as follows:
\begin{equation}
    y_k^R = h_k^R \sqrt{p^C} s_0 + \jsum h_k^R \sqrt{p_k^P} s_j + n,
\end{equation}
where $h^R_k$ is the corresponding sub-channel gain of users operating in RSMA, $s_0$ is the common message, and $p^C$ and $p_k^P$ are the transmit powers of the common signal and private signals, respectively.
For RSMA, we have two data rates, common and private rates, and the total rate for a user is the summation of these two rates. The achievable common rate of user $k$ can be stated as follows: 
\begin{equation}\label{eq:CommonRate}
    R_k^C = \log_2(1+{\gamma}_k^C)= \log_2\left(1+\frac{\abss{h^R_k}p^C}{\abss{h^R_k}\jsum p_j^P + \sigma^2}\right),
\end{equation}
where $\gamma_k^C$ denotes the common SINR of user $k$ in RSMA.
In \eqref{eq:CommonRate}, the private signals of all users, $\abss{h^R_k} \jsum p_j^P$, are considered as \textcolor{black}{noise} when user $k$ is decoding the common signal. To ensure that all users can decode the common signal, the common rate must not exceed $R^C = \min_{k=1,...,U} \ R_{k}^C$. Also, by defining $c_k$ as the common rate of user $k$, we need to ensure that the maximum common rate that can be sent to users, i.e., $R^C$ is shared between all users, hence, $\ksum c_k = R^C$ \cite{CommRateConstraint,SICforRSMA}. 
Note that SIC is performed only once in the RSMA scheme when subtracting off the common signal to decode the private one in each user. Therefore, the SIC constraint only requires the difference between the common signal power and the total private signals powers to be greater than the SIC threshold, i.e., $P_{\mathrm{tol}}$ \cite{SICforRSMA}. Therefore we have the following constraint on the common and private powers.
\begin{equation}\label{eqn:SIC-RSMA}
    p^C \delta^R_k - \jsum p^P_j \delta^R_k \geq P_{\mathrm{tol}}, \forall k \in \Uset,
\end{equation}
where $\delta^R_k = \frac{\abss{h^R_k}}{\sigma^2}$. Considering that the other users' private signals are treated as noise, the private rate of user $k$ can be expressed as follows:
\begin{equation}
    R_k^P =  \log_2(1+{\gamma}_k^P) = \log_2\left(1+\frac{\abss{h^R_k}p_k^P}{\abss{h^R_k}\jNotKsum p_j^P + \sigma^2}\right).
\end{equation}
Therefore, the total rate of user $k$ in RSMA is given by $$R_k^R=c_k+R_k^P.$$
As a result of using two frequencies at each user, each of which works at either NOMA or RSMA, the total rate of each user is stated as follows.
\begin{equation}\label{eqn:Rate}
    R_k =  R_k^N + R_k^P + c_k.
\end{equation}



\section{Weighted Sum-Rate Maximization in Hybrid RSMA-NOMA}
In order to utilize the advantages of both NOMA and RSMA, we propose a hybrid access scheme that benefits from both access methods. By introducing distinct weights for each access method, we can take advantage of RSMA's greater achievable rate for nearby users and NOMA's higher rate for distant users. To do so, we aim to maximize the weighted sum rate in a hybrid RSMA-NOMA system in the following optimization problem.  

\begin{maxi!}[3]
	{\substack{c_k, \beta^N, p_k^N\\ p^C, p_k^P, \forall k \in \Uset}}
	{\ksum \omegaN R_k^{N} + \ksum \omegaR (R_k^{P} + c_k)\label{eq:P1Obj}}
	{\label{problem:p1main}}{\mathcal{P}_1: \ \ }
	\addConstraint{\isum  c_i}{\leq R^C_{k} \ \forall k \in \Uset \label{eq:p1RMSA1}}
	\addConstraint{R_k^{N} + R_k^{P} + c_k }{\geq R_k^{\mathrm{th}} \ \forall k \in \Uset \label{eq:p1QOS}}
	\addConstraint{\ksum  p_k^N}{\leq \beta P_{\max} \label{eq:p1MaxPowerNOMA}}
	\addConstraint{\ksum p_k^P + p^C}{\leq (1-\beta)P_{\max} \label{eq:p1MaxPowerRSMA}}
	\addConstraint{p_k^N \delta^N_{k-1} - \kMOnesum p_j^N \delta^N_{k-1}}{\geq P_{\mathrm{tol}}, \ \forall k=2,3,...,U \label{eq:p1SICNOMA}}
	\addConstraint{p^C \delta^R_k - \jsum p^P_j \delta^R_k \geq P_{\mathrm{tol}}, \forall k \in \Uset \label{eq:p1SICRSMA}}
\end{maxi!}
where \eqref{eq:p1RMSA1} guarantees that the maximum total common rate, $R_C$, is distributed among all users. \eqref{eq:p1QOS} is the quality of service (QoS) constraint for user $k$ and $R_k^{th}$ is the minimum rate requirement for user $k$. \eqref{eq:p1MaxPowerNOMA} and \eqref{eq:p1MaxPowerRSMA} are maximum transmission power constraint, which also include the power budget allocation between RSMA and NOMA channels. \eqref{eq:p1SICNOMA} and \eqref{eq:p1SICRSMA} are SIC constraints for NOMA and RSMA, respectively.
The optimization problem $\pr_1$ is non-convex due to the coupled optimization variables and the non-convex objective function and constraints \eqref{eq:p1RMSA1}, \eqref{eq:p1QOS}, therefore it cannot be solved directly. To solve the non-convex problem, we employ the difference of convex (DC) and SCA methods that approximate the non-convex objective function and constraints with a convex function using a linear approximation. Then the resulting convex problem is solved iteratively to obtain resource allocation solution. 

\section{Resource Allocation}

This section introduces an iterative solution to the non-convex problem $\pr_1$ to jointly optimize users' powers, rate, and power budget allocation.

\subsection{Joint Power and Rate Allocation}
{In this subsection, $\pr_1$ is solved to obtain joint power and rate allocation for both NOMA and RSMA schemes using SCA and DC methods. Finally, the approximated convex problem is solved iteratively.}
Due to the objective function and constraints \eqref{eq:p1RMSA1}, \eqref{eq:p1QOS}, $\pr_1$ is non-convex. By leveraging the logarithm properties, the objective function \eqref{eq:P1Obj} can be rewritten as follows:
\begin{equation} \label{eq:objLog}
    \begin{split}
        f(\boldsymbol{P,C})&\!=\! \ksum \omegaN R_k^{N,D} \!\!-\!\! \ksum \omegaN R_k^{N,I} \!+\! \ksum \omegaR R_k^{R,D} \\ &-\!\!\ksum \omegaR R_k^{R,I}\!+\!\!\ksum \omegaR c_k,
    \end{split}
\end{equation}
where $R_k^{N,D} = \log(\sum\limits_{j=1 }^{k} p_j^N + a_k^N)$, $R_k^{N,I} = \log(\kMOnesum p_j^N + a_k^N)$, $R_k^{R,D} = \log(\jsum p_j^P+a_k^R)$, $R_k^{R,I} = \log(\jNotKsum p_j^P+a_k^R)$, such that $a_k^i =\frac{\sigma^2}{\abss{h_k^i}}, \forall i \in \{N,R\}$, and $\boldsymbol{P}=[p_k^N,p_k^P,p^C]$, $\boldsymbol{C}=[c_k], \ \forall k \in \Uset$.

It can be shown that $R_k^{N,D}$ and $R_k^{R,D}$ are concave functions of $\boldsymbol{(P,C)}$. Using the first-order Taylor approximation of these two convex functions, we can maximize a lower bound of the objective function iteratively, thereby maximizing the original function \cite{boyd2004convex}. Therefore, the lower bound of the objective function is stated as follows:
\begin{equation*}
    \hat{f}(\boldsymbol{P,C})\!=\!\!\! \ksum\!\!\omegaN R_k^{N,D}\!-\!\!\!\ksum\!\omegaN\log(\!\kMOnesum p_j^{N}(t) \!+\! a_k^N) \!
\end{equation*}
\begin{equation*}
 - \!\!\!\!  \sum\limits_{k=1 }^{U-1} \!\!\left(\!\!\sum\limits_{i=k+1 }^{U}\frac{\omega_i^{N}}{\sum\limits_{j=1 }^{i-1}p_j^{N}(t)\!+\!a_i^N}\!\right)\!\!\frac{(p_k^N\!-\!p_k^{N}(t))}{\mathrm{ln}(2)}
\end{equation*}    
\begin{equation*}
    \hspace{-12mm}  +\!\!\! \ksum \omegaR \! R_k^{R,D}\! - \!\!\!\!\ksum \omegaR\!\log(\jNotKsum\! p_j^{P}(t)\!\!+\!a_k^R)\!\!    
\end{equation*}
 \begin{equation}\label{eq:PowerAllObjFunc}
     -\!\!\ksum\! \left(\!\sum\limits_{\substack{i=1 \\ i\neq k}}^U  \!\frac{\omega_i^{R}}{\sum\limits_{\substack{j=1 \\ j\neq i}}^U p_j^{P}(t)\!+\! a_i^R} \!\!\right)\!\frac{(p_k^P\!\!-\!p_k^{P}(t))}{\mathrm{ln}(2)}\! +\!\!\ksum \omegaR c_k,
\end{equation}
where $x(t)$ is used to denote the value of variable $x$ at $t$-th iteration.
Similarly, we can apply the same approach so as to handle the non-convexity of QoS constraints in \eqref{eq:p1QOS} and rewrite them as follows.
\begin{equation}\label{eq:PowerAllQoSAppr}
     R_k^{N,D}\!\!\!\!-\!\log(\!\kMOnesum \!p_j^{N}(t)\!\!+\!a_k^N)\! -\! \frac{\!\!\kMOnesum \!(p_j^N\!-\!p_j^{N}(t))}{\!(\!\kMOnesum\! p_j^{N}(t)\!\!+\!a_k^N)\mathrm{ln}(2)}  \nonumber
\end{equation}
\begin{equation}\label{eq:PowerAllQoSAppr-RSMA}
    \begin{split}
        & + R_k^{R,D} \!\!-\log(\jNotKsum p_j^{P}(t)\!\!+a_k^R) \!-\! \frac{ \jNotKsum (p_j^P-p_j^{P}(t))}{(\jNotKsum p_j^{P}(t)+a_k^R)\mathrm{ln}(2)} \\ &+ c_k \geq R_k^{\mathrm{th}}, \ \forall k \in \Uset
    \end{split}
\end{equation}

With the purpose of handling the non-convexity of constraint \eqref{eq:p1RMSA1}, we can introduce slack variable $\gamma_k $ and rewrite the constraint as follows.
\begin{subequations}
\begin{align}
\ksum c_k \leq \log(1+\gamma_k), \ \forall k \in \Uset \label{eq:RSMAconsTgamma1}\\
\gamma_k \leq \frac{p^C}{\jsum p_j^P +a^R_k}, \ \forall k \in \Uset \label{eq:RSMAconsTgamma2}
\end{align}
\end{subequations}
Constraint \eqref{eq:RSMAconsTgamma2} is non-convex, so by adding slack variable $\lambda_k$, it can be replaced by the following constraints.
\begin{subequations}
\begin{align}
& \lambda_k \geq \jsum p_j^P + a^R_k, \ \forall k \in \Uset \label{eq:RSMAconsTLambda1}\\
& p^C \geq \lambda_k\gamma_k = \frac{1}{4}({(\lambda_k+\gamma_k)}^2 - {(\lambda_k-\gamma_k)}^2), \ \forall k \in \Uset \label{eq:RSMAconsTLAmbda2}
\end{align}
\end{subequations}
Right-hand side of \eqref{eq:RSMAconsTLAmbda2} is difference of convex and can be approximated to a convex function using first-order Taylor approximation. Thus, \eqref{eq:RSMAconsTLAmbda2} can be given as follows:
\begin{equation}\label{eq:RSMAconsTlambda2Appr}
\begin{split}
    p^C \geq & \frac{1}{4}({(\lambda_k+\gamma_k)}^2 + {(\lambda_k(t)-\gamma_k(t))}^2 \\&- 2(\lambda_k(t)-\gamma_k(t))(\lambda_k-\gamma_k)), \ \forall k \in \Uset.  
\end{split}
\end{equation}
Considering the above approximations, $\pr_1$ can be given as: 
\begin{maxi!}[2]
	{\substack{p_k^N, p^C, p_k^P,\gamma_k,\lambda_k \\ c_k, \beta^N, \ \forall k \in \Uset}}
	{\hat{f}(\boldsymbol{P,C})\label{eq:P3Obj}}
	{\label{problem:p3main}}{\mathcal{P}_2: \ \ }
    \addConstraint{\eqref{eq:RSMAconsTgamma1},\eqref{eq:RSMAconsTLambda1},\eqref{eq:RSMAconsTlambda2Appr},\eqref{eq:PowerAllQoSAppr}\nonumber}
    \addConstraint{\eqref{eq:p1MaxPowerNOMA},\eqref{eq:p1MaxPowerRSMA},\eqref{eq:p1SICNOMA},\eqref{eq:p1SICRSMA} \nonumber}
\end{maxi!}
Problem $\pr_2$ at iteration $(t)$ is a convex optimization problem that can be solved efficiently using standard solvers such as CVX \cite{boyd2004convex}.
Finally, the overall procedure to solve $\pr_1$ is summarized in algorithm 1. 
\begin{algorithm}[t]
\small
    \caption{Weighted Sum-rate Maximization: Hybrid RSMA-NOMA}
    \label{Algorithm 1}
    \textbf{Input}: Initial values of variables, RSMA and NOMA users weights $\omegaN$ and $\omegaR$, Maximum number of iterations $L_{\max}$, stopping accuracy $\epsilon_1$.
    \begin{algorithmic}[1]
        \For {$t=0,1,...$}
        \State Solve $\pr_2$ and obtain $p_k^{N}(t), p_k^{P}(t), p^C, c_k, \beta$
        \State Set $t = t+1$.
        \State \textbf{Until} $\hat{f}(\boldsymbol{P,C})^{(t+1)} - \hat{f}(\boldsymbol{P,C})^{(t)} < \epsilon_1$ or $t=L_{\max}$
        \EndFor
    \end{algorithmic}
    \textbf{Output}: The optimal solutions: $p_k^{N,*}, p_k^{P,*}, p^{C,*}, c_k^{*}, \beta^{*}, R_k^{*} \forall k \in \Uset$.
\end{algorithm}
\subsection{Complexity Analysis}

This section provides a complexity analysis of algorithm 1. Solving the approximated convex optimization problem $\pr_1$ requires the complexity of $\mathcal{O}({(5U+2)}^2 (6U+1))$, which is based on $\mathcal{O}(Q_1^2Q_2)$, where $Q_1$ and $Q_2$ are the number of variables and constraints of $\pr_2$, respectively \cite{lobo1998applications}. Moreover, the number of required iterations for the SCA method used to solve $\pr_1$ is $\mathcal{O}(\sqrt{4U+1}\log(\frac{1}{\epsilon_1}))$, where $4U+1$ is the number of constraints in $\pr_1$ and $\epsilon_1$ is the stopping accuracy \cite{cvx}. Therefore, the total complexity of Algorithm 1 can be written as $\mathcal{O}(U^{3.5}\log(\frac{1}{\epsilon_1}))$.

\begin{figure}[!t]
   \centering
\includegraphics[scale=0.57]{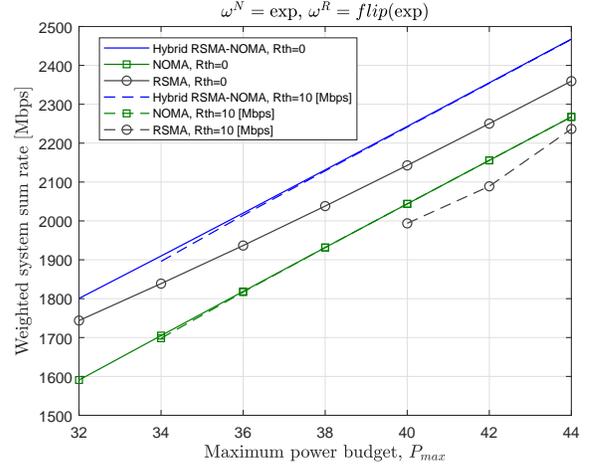}
\caption{Weighted system sum-rate with unequal weights (objective function of $\pr_1$)}
\label{fig:WSSR}
\end{figure}

\begin{figure*}[t]
\begin{minipage}{0.48\textwidth}
\includegraphics[scale=0.57]{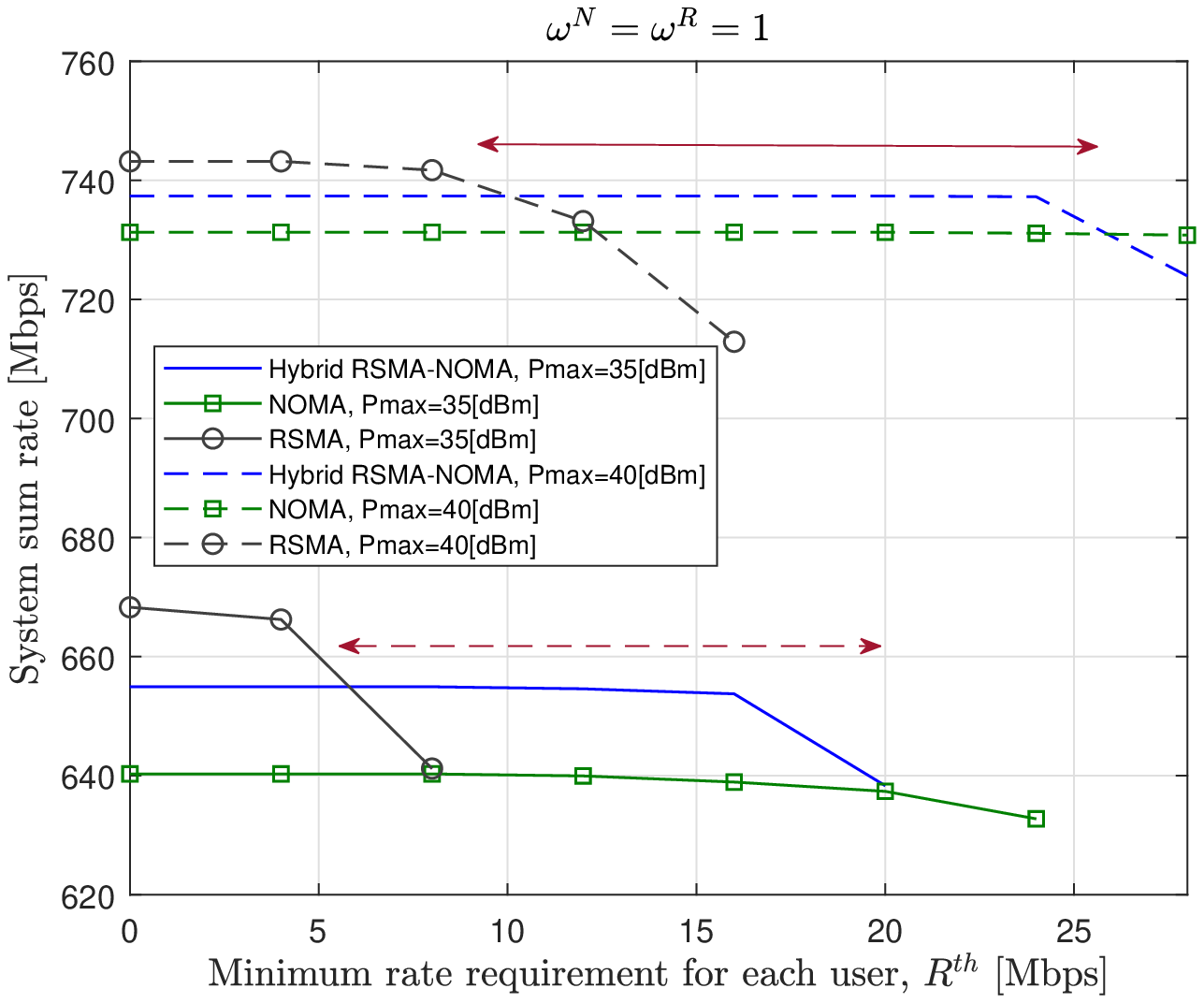}
\caption{System sum-rate versus the minimum required rate, $R^{\mathrm{th}}$, with equal users' weights set to unity.}
\label{fig:SSR_Rth_Change_WeightsOnes}
\end{minipage}\hfill
\begin{minipage}{0.48\textwidth}
\includegraphics[scale=0.57]{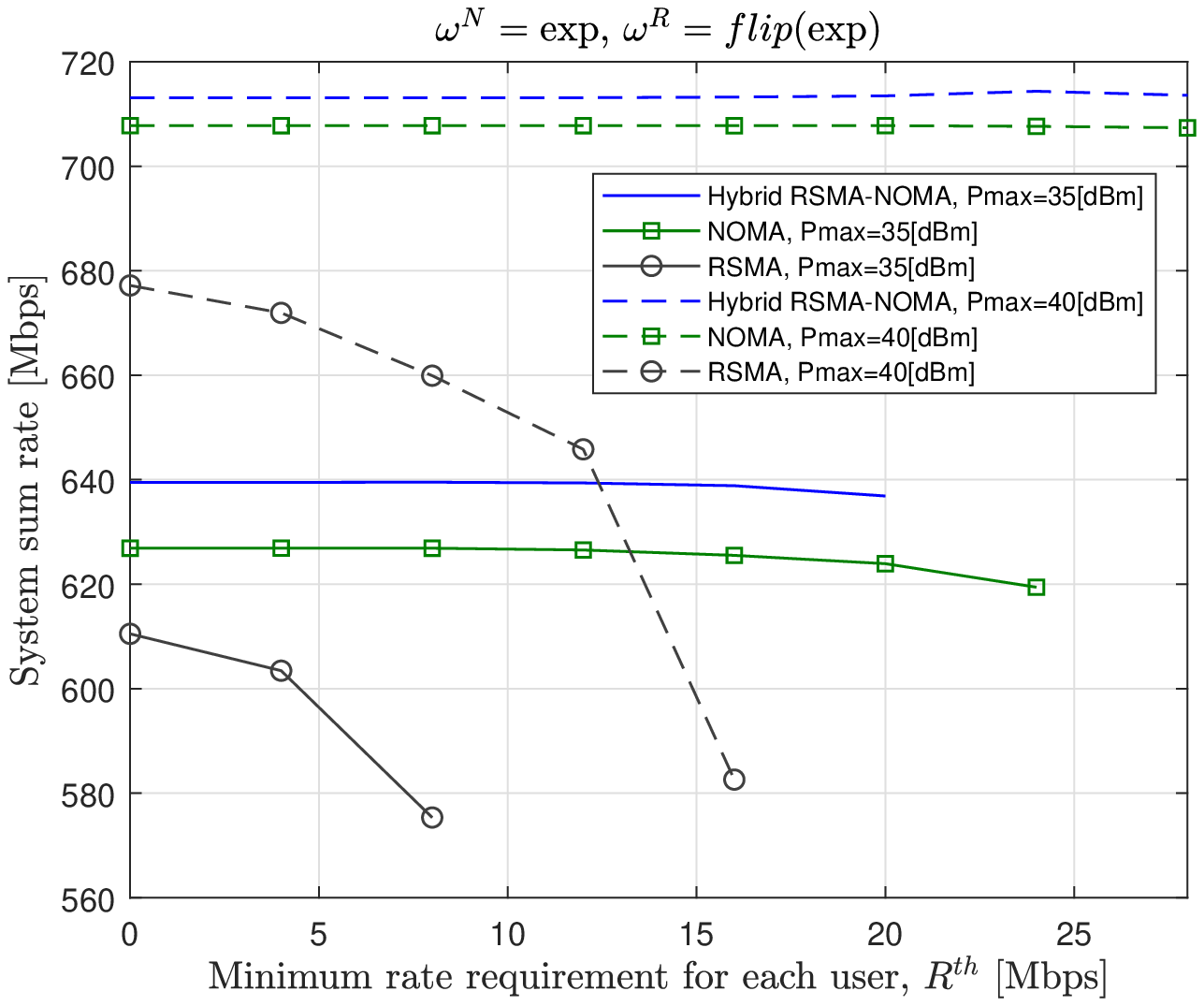}
\caption{System sum rate versus the minimum required rate, $R^{\mathrm{th}}$, with distinct weights for NOMA and RSMA}
\label{fig:SSR_RthChange_WeightsExp}
\end{minipage}
\end{figure*}

\begin{figure*}[t]
\begin{minipage}{0.48\textwidth}
\includegraphics[scale=0.57]{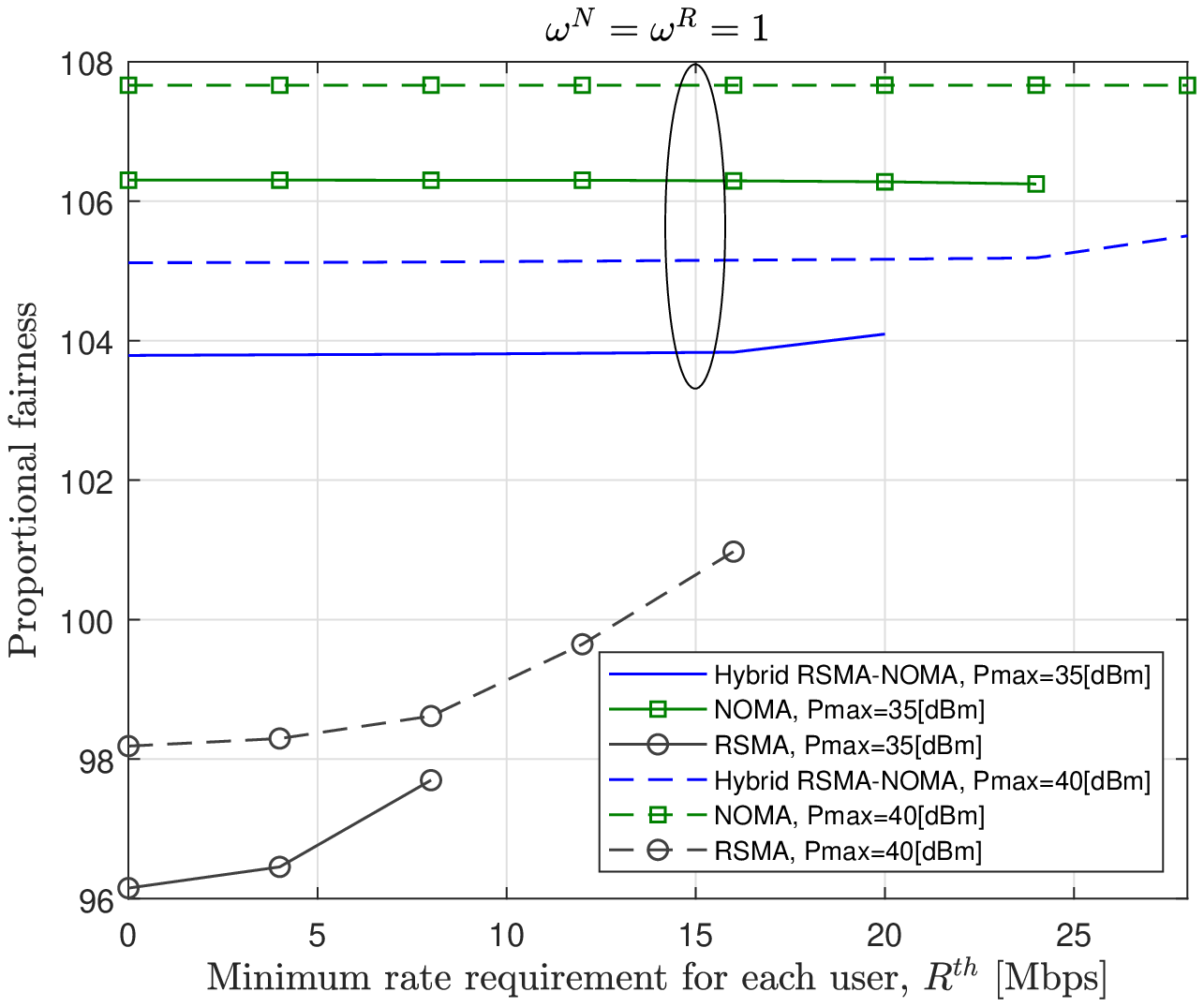}
\caption{Proportional fairness versus the minimum required rate, $R^{\mathrm{th}}$, with weights equal to one}
\label{fig:ProporFairness_Rth_Change_WeightsOnes}
\end{minipage}\hfill
\begin{minipage}{0.48\textwidth}
\includegraphics[scale=0.57]{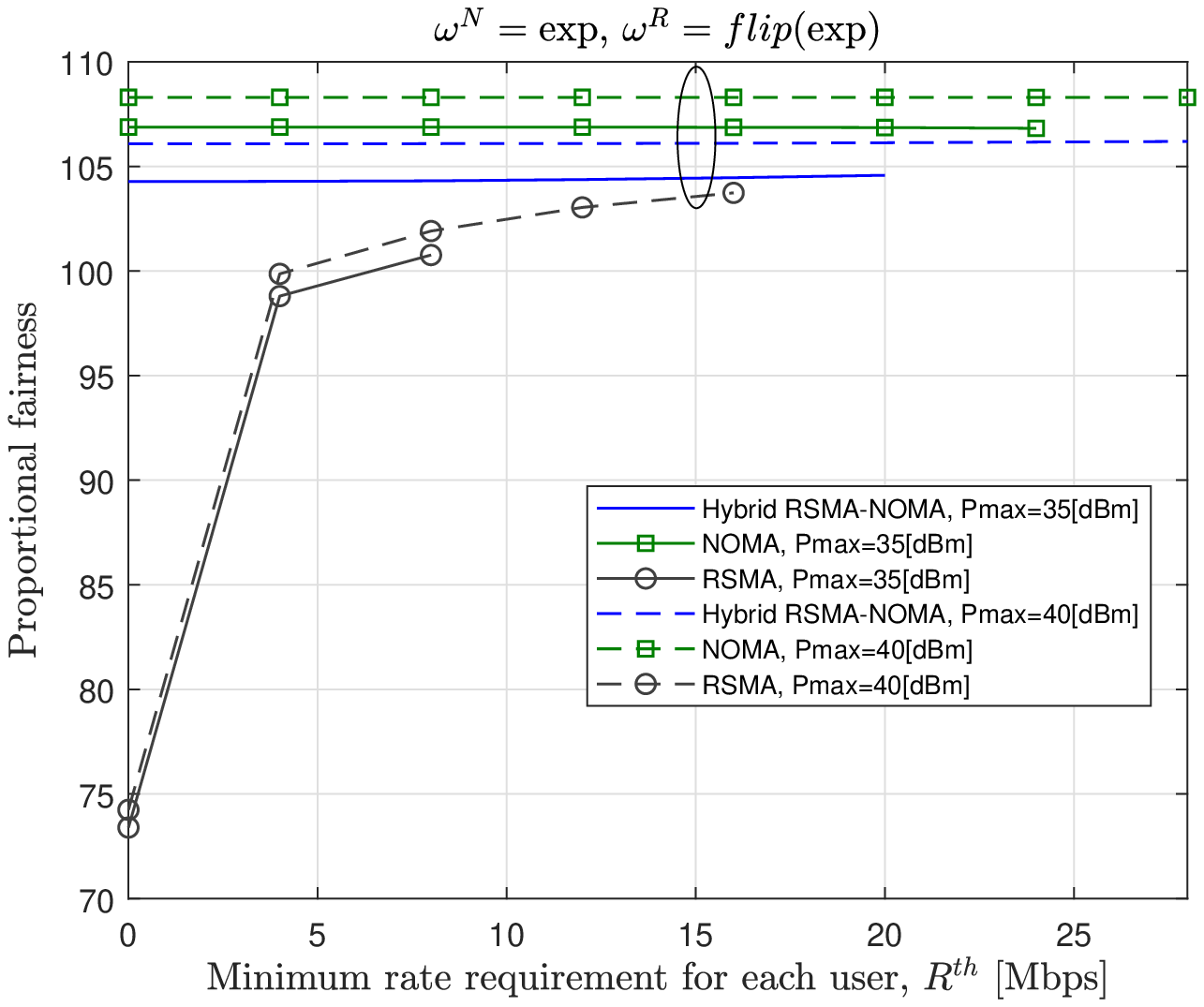}
\caption{Proportional fairness versus the minimum required rate, $R^{\mathrm{th}}$, with distinct weights for NOMA and RSMA}
\label{fig:ProporFairness_RthChange_WeightsExp}
\end{minipage}
\end{figure*}
\section{Numerical Results}
In this section, several numerical results are provided to evaluate the performance of the proposed hybrid NOMA-RSMA access scheme and the impact of selecting different weights for each access scheme in hybrid RSMA-NOMA. It is assumed that the BS is located at (0,0,4) [m], and users are distributed over a $350\times350$ [m$^2$] area. Similar to \cite{SICforRSMA}, the large-scale path loss is modeled by $\text{PL} = -128.1-37.6\log_{10}(d)$, where $d$ denotes the distance between two nodes in [km] \cite{PathLossModel}. All links are assumed to be non-line-of-sight (NLOS) which are modeled by Rayleigh fading distribution. Other simulation parameters are specified in Table 1. In what follows, \textit{flip()} denotes the reverse operation, that is when $\omegaN=\exp(0.24k)$, RSMA weights are $\omegaR = \exp(0.24(U-k))$.

\begin{table}[thb!]
\caption{Numerical Results Parameters}
\label{tab:my-table}
\centering
\begin{tabular}{|c|c|}
\hline
\textbf{Parameter}                                            & \textbf{Value}    \\ \hline
Maximum transmission power at BS, $P_{\max}$                  & 35 {[}dBm{]}      \\ \hline
Noise power at users                                          & -110 {[}dBm{]}    \\ \hline
Algorithm convergence parameters, $\{\epsilon_1, L_{\max}\}$ & $\{10^{-3},100\}$ \\ \hline
SIC detection threshold, $P_{\mathrm{tol}}$ & $10 \  \text{[dBm]}$ \\ \hline
\end{tabular}
\end{table}

The objective function in $\pr_1$ that must be maximized is the weighted system sum rate, as depicted in Figure \ref{fig:WSSR}. This figure demonstrates that when the maximum power budget increases, selecting various weights for each access scheme in the hybrid RSMA-NOMA can improve the weighted system sum rate compared to NOMA and RSMA. {It is also evident that RSMA renders infeasibility with a smaller maximum power budget compared to hybrid scheme and NOMA. Also, the degree of degradation of RSMA is observed to be much higher with the increase in minimum rate requirements of all users $R^\mathrm{th}$.}


\subsection{System Sum-rate Analysis and Insights}
Figure \ref{fig:SSR_Rth_Change_WeightsOnes} illustrates system sum rate when the minimum required rate for each user, i.e., $R^{\mathrm{th}}$, varies and all weights are set to one. It can be observed that as $R^{\mathrm{th}}$ increases, the data rate of all schemes begin to decrease, as a result of restricting the constraint set of the optimization problem $\pr_1$. Although RSMA initially provides a higher system sum rate, its faster rate degradation with $R^{\mathrm{th}}$   makes it less efficient compared to other access methods. On the other hand, it can be seen that the hybrid RSMA-NOMA outperforms NOMA and RSMA after a certain point, even with a lower maximum power budget. Also, as the maximum power increases, the region where hybrid RSMA-NOMA prevails expands. In Figure \ref{fig:SSR_RthChange_WeightsExp}, the effect of increasing $R^{\mathrm{th}}$ is depicted using distinct weights. Compared to Figure \ref{fig:SSR_Rth_Change_WeightsOnes}, it is evident that the proposed hybrid RSMA-NOMA consistently outperforms NOMA and RSMA. Particularly when a larger power budget is used, hybrid RSMA-NOMA does not lie within the infeasible region and behaves similarly to NOMA in terms of feasibility. 

\subsection{Fairness Analysis and Insights}
{Figure \ref{fig:ProporFairness_RthChange_WeightsExp} demonstrates proportional fairness when scenario \textbf{(ii)} is employed, i.e., distant users are prioritized more in NOMA, whereas nearby users are prioritized more in RSMA.}
Compared to scenario \textbf{(i)}, which is depicted in Figure \ref{fig:ProporFairness_Rth_Change_WeightsOnes}, it can be seen that proportional fairness has improved. It can be concluded that by employing various weights in the proposed hybrid RSMA-NOMA scheme, greater system fairness can be achieved at the expense of a slight reduction in system sum rate. 
In order to compare the impact of each access method on fairness, we use proportional fairness metric, denoting the sum of logarithmic average of data rates of users, which can be written as $\ksum \mathrm{ln}(R_k)$ \cite{ProportionalFairness}.  It can be seen that fairness improves by the minimum rate. Specifically, fairness enhances in hybrid RSMA-NOMA and RSMA, whereas it remains unchanged in NOMA.

\begin{figure}[!t]
   \centering
\includegraphics[scale=0.57]{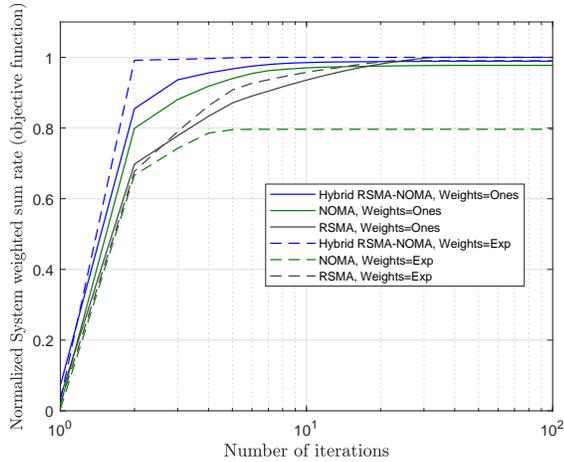}
\caption{Comparison of convergence}
\label{fig:Convergence}
\end{figure}
\subsection{Convergence}
Figure \ref{fig:Convergence} illustrates the differences in the access schemes with regards to algorithm convergence, which is a function of the complexity of the algorithms. It is clear to see that the hybrid RSMA-NOMA converges faster than both the RSMA and the NOMA algorithm when using $\omegaN=\omegaR=1$. Additionally, RSMA has the slowest rate of convergence, which can be rationalized by considering the higher number of convex approximations that are necessary to be considered at each iteration of SCA when RSMA is utilized. These approximations arise from the common rate constraints that are placed on the system. On the other hand, by choosing different weights for the hybrid RSMA-NOMA, one can drastically improve the convergence rate, which also helps to improve the convergence rate of the RSMA.

\section{Conclusion}
In this paper, we propose a hybrid RSMA-NOMA access method in which the BS transmits user signals employing RSMA and NOMA over two orthogonal sub-channels. Power and rate allocation are considered in order to maximize the weighted system sum rate, while practical constraints, including QoS and SIC constraints, and the proportion of the maximum power budget for each access scheme are taken into account. In order to deal with the non-convexity of the optimization problem, a low-complexity SCA algorithm is proposed. In the hybrid RSMA-NOMA method, system sum rate, and proportional fairness can be enhanced by utilizing the capability of RSMA to deliver a higher data rate and NOMA to provide greater fairness. Particularly, when a distinct set of weights is designated to each access scheme, fairness is improved while system sum rate is kept as high as possible. It is also demonstrated that the proposed hybrid RSMA-NOMA outperforms RSMA and NOMA with equally assigned weights when all users increase their minimum required rate.
\bibliography{ref}
\bibliographystyle{IEEEtran}

\end{document}